\documentclass[11pt]{article}
\usepackage{graphicx}

\title{Low-lying Dirac eigenmodes and monopoles in 4D compact QED}
\author{Toru T. Takahashi
\thanks{Yukawa Institute for Theoretical Physics, Kyoto university,
Kitashirakawa-Oiwakecho, Sakyo, Kyoto 606-8502, Japan}}

\begin{document}

\maketitle

\abstract{
We study the properties of low-lying Dirac modes
in quenched compact QED at $\beta$=0.99, 1.01 and 1.03,
employing $12^3\times 12$ lattices and the overlap formalism for the
fermion action. 
We pay special attention to the spatial distributions
of the low-lying Dirac modes.
Near-zero modes are found to have universal anti-correlations
with monopole currents below/above the critical $\beta$.
We also study the nearest-neighbor level spacing distribution
of Dirac eigenvalues and find a signal of a Wigner-Poisson transition.
We make a possible speculation on the chiral phase transition
in 4D compact QED.
}

\section{Introduction}
\label{introduction}

The chiral symmetry breaking 
is one of the most famous nonperturbative phenomena in QCD,
which have been attracting a great deal of interest for a long time.
Chiral symmetry, which is an approximate global symmetry in QCD,
is spontaneously broken by nonperturbative dynamics of QCD.
The broken chiral symmetry leads to large constituent quark masses
of a few hundred MeV~\cite{DeRujula:1975ge,Nambu:1961tp,Nambu:1961fr}, 
which are responsible for about 99\% of mass in the world,
aside from some unknown factors such as dark matters.

The spontaneous chiral symmetry breaking
is caused by non-zero chiral condensate
$\langle \bar \psi \psi \rangle$,
which is directly related to non-vanishing spectral density
at the spectral origin of the Dirac operator,
via the Banks-Casher relation~\cite{Banks:1979yr}.
The level dynamics of Dirac eigenvalues
is then essential for the chiral phase transition.
As for QCD,
the level statistics of the low-lying eigenvalues
of the Dirac operator has been studied with much effort and 
it is known to be reproduced by the random matrix theory (RMT)
with the same global symmetry.
There however remain several issues to be studied
from the microscopic viewpoint.

The spontaneous chiral symmetry breaking
can be also seen in compact QED in 1+3 dimensions.
Compact QED in a strong-coupling region exhibits the charge confinement
as well as the spontaneous chiral symmetry breaking like QCD.
Its nature has been also studied by many groups so far
~\cite{DeGrand:1980eq,Barber:1984ak,Grosch:1985cz,Barber:1985at,
Kogut:1987kg,Azcoiti:1987ji,Dagotto:1987ik,Azcoiti:1989ue,
Salmhofer:1990xs,Hashimoto:1990nh,Gausterer:1991ef,Azcoiti:1991ng,
Azcoiti:1994ku,Baig:1994ib,Shiba:1994pu,Bielefeld:1997nh,Cox:1997ek,
Berg:1998xv,Berg:2001nn,Arnold:2002jk,Drescher:2004st,
Vettorazzo:2004cr,Panero:2005iu}.

The key ingredient 
for the nonperturbative features in compact QED would be monopoles
in a microscopic sense.
Monopoles play essential roles in the charge confinement and its
phase transition, and would play important roles 
also in the chiral symmetry breaking.
As a matter of fact,
neither the confinement nor the chiral symmetry breaking occurs
without monopoles' degrees of freedom.
In this paper,
we mainly pay attention to the spatial distributions of low-lying Dirac
eigenmodes in quenched compact QED,
because the low-lying eigenvalue dynamics,
which is responsible for the chiral symmetry breaking,
is considered to have a close connection with the low-lying eigenfunctions.
We investigate
the spatial correlations between low-lying eigenfunctions and
monopoles, the key player in the nonperturbative dynamics in compact QED.

The paper is organized as follows.
We give the formalism employed in the present analysis in Sec.~\ref{formalism}.
In Sec.~\ref{diracmodes}, the fundamental features of the low-lying
Dirac modes are summed up.
We show the lattice QED results in Sec.~\ref{analysis}.
Sec.~\ref{discussions} is devoted to the discussions and the
speculations based on the present lattice QED results.
We finally make a summary in Sec.~\ref{summary}.

\section{Formalism}
\label{formalism}

\subsection{compact QED}

The Wilson gauge action for compact QED is written as
\begin{equation}
S_{\rm QED}
=
\beta \sum_{x}\sum_{\mu , \nu}(1-{\rm Re\ }P_{\mu\nu}(x))
\end{equation}
with link variables
$U_\mu(x)\equiv e^{i\theta_\mu(x)}\in U(1)$
and plaquettes
$P_{\mu\nu}(x)\equiv U_\mu(x) U_\nu(x+\hat\mu) U^\dagger_\mu(x+\hat\nu)U^\dagger_\nu(x)$.
The constant $\beta\equiv \frac{1}{e^2}$ corresponds to the coupling constant.
Here, the angle $\theta_\mu(x)$ ranges from $-\pi$ to $\pi$.
Defining a plaquette angle $\theta_{\mu\nu}(x)
\equiv \theta_\mu(x)+\theta_\nu(x+\hat\mu)-\theta_\mu(x+\hat\nu)-\theta_\nu(x)
\in (-4\pi , 4\pi]$, $S_{\rm QED}$ is represented as
\begin{equation}
S_{\rm QED}
=
\beta \sum_{x}\sum_{\mu , \nu}(1-\cos\theta_{\mu\nu}(x)).
\end{equation}
This compact formulation leads to several nonperturbative phenomena,
such as the chiral symmetry breaking or the charge confinement at small $\beta$.
Indeed, at zero temperature, compact QED has two phases;
the confinement phase and the Coulomb phase.
These two phases are separated with the critical value $\beta_c = 1.0111331(21)$~\cite{Arnold:2002jk}.
The system is in the confined phase at $\beta < \beta_c$,
while it's in the Coulomb phase at $\beta > \beta_c$.
The phase transition is weak first order.

In this paper,
we employ $12^3\times 12$ lattices at $\beta$=0.99, 1.01 and 1.03.
We generate and investigate 
independent 48 gauge configurations at each $\beta$,
which are generated with the standard Wilson gauge action imposing
the periodic boundary conditions in all the directions.
The system at $\beta$=0.99 (1.03) clearly lies in 
the confinement (Coulomb) phase, respectively,
but the $12^3\times 12$ system at $\beta$=1.01 is marginal.
``Finite temperature'' phase transition in compact QED
was extensively investigated in Ref.~\cite{Vettorazzo:2004cr},
and the ``transition temperature'' $1/T$ at $\beta$=1.01
is found to be about $1/T \sim 6$.
Then the $12^3\times 12$ system at $\beta$=1.01 is considered to be
in the confinement phase.

As we mentioned in Sec.~\ref{introduction},
the key ingredient in compact QED in the confinement phase
is monopoles' degrees of freedom.
In order to extract monopoles,
we divide plaquette angles $\theta_{\mu\nu}$ into two parts;
physical fluxes $\bar\theta_{\mu\nu}\in (-\pi,\pi]$ 
and Dirac strings $2\pi n_{\mu\nu}$.
\begin{equation}
\theta_{\mu\nu} = \bar\theta_{\mu\nu} + 2\pi n_{\mu\nu}
\end{equation}
$n_{\mu\nu}\in [0,\pm 1,\pm 2]$
is integer-valued and corresponds to the number of
the Dirac strings penetrating the plaquette.
We can now define the integer-valued 
DeGrand-Toussaint monopole current $m_\mu (x)$~\cite{DeGrand:1980eq}
in a gauge-invariant manner;
\begin{equation}
m_\mu(x) = \frac12 \varepsilon_{\mu\nu\kappa\lambda}
\Delta_\nu^+n_{\kappa\lambda}(x),
\end{equation}
where $\Delta_\mu^\pm$ is a forward and backward derivative
operator on a lattice, respectively.
The monopole currents $m_\mu(x)$ satisfy the conservation law
$\Delta_\mu^- m_\mu(x)=0$ and hence form closed loops.

Monopole currents can be unambiguously classified
into monopole clusters $C_{\rm mon}^i$. 
In Ref.~\cite{Hart:1997vb}, it was conjectured that 
the largest monopole cluster $C_{\rm mon}^1$
occupies most of monopole currents
and only the largest monopole cluster is relevant for the color confinement.
\begin{figure}[h]
\begin{center}
\includegraphics[scale=0.27]{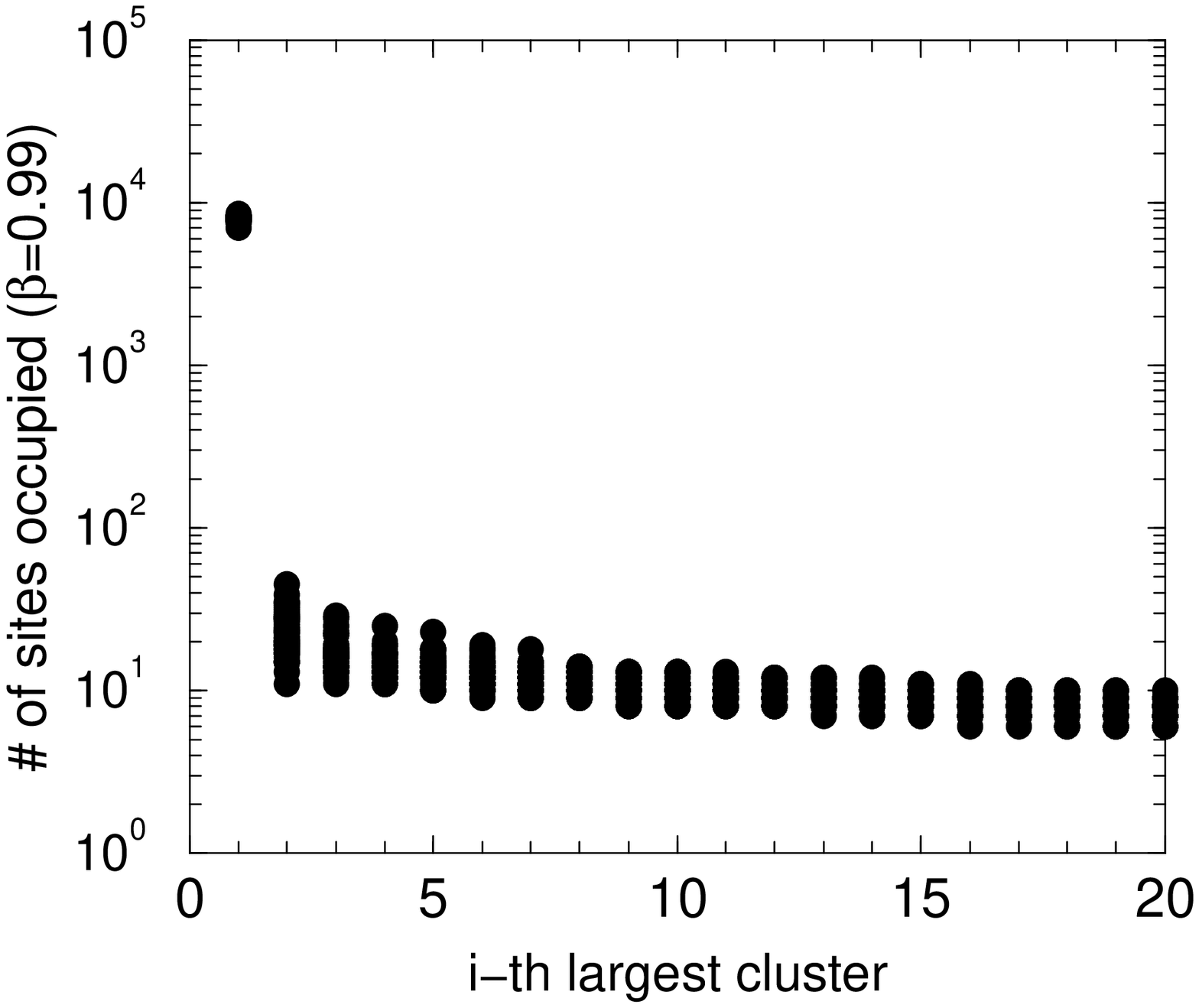}
\includegraphics[scale=0.27]{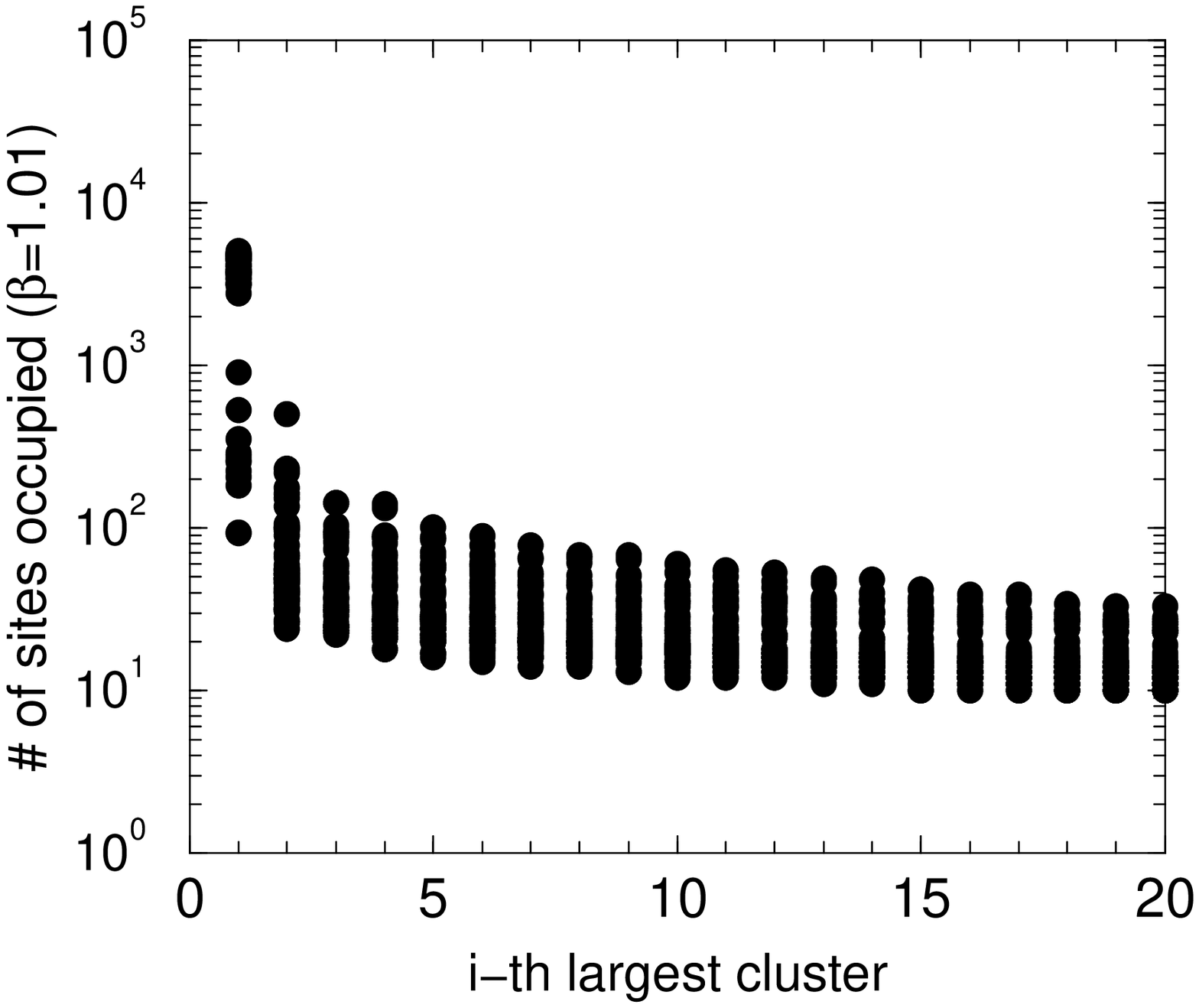}
\includegraphics[scale=0.27]{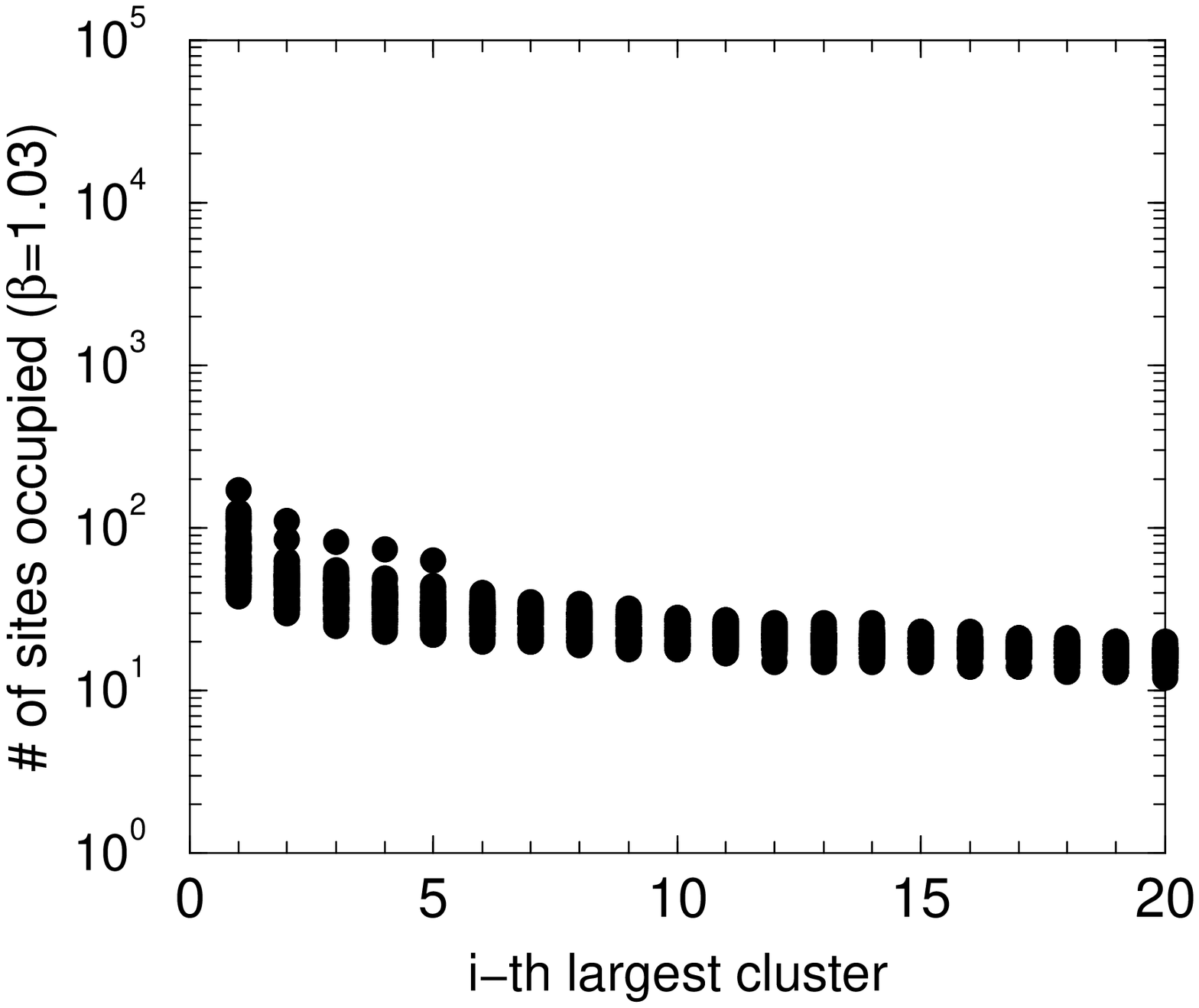}
\end{center}
\caption{\label{mch01}
The numbers of the sites occupied by the $i$-th largest monopole cluster
are plotted, which are obtained with 48 configurations at each $\beta$. 
}
\end{figure}
In Fig.~\ref{mch01}, we show the scattered plots of 
the numbers of the sites occupied by the $i$-th largest
monopole cluster at each $\beta$,
which are obtained with 48 configurations at each $\beta$.
At $\beta$=0.99 and 1.01,
we find prominently large monopole clusters
of the length of about 10000,
whereas we find no large cluster at $\beta$=1.03.
The second largest cluster $C_{\rm mon}^2$
is much smaller than the largest cluster $C_{\rm mon}^1$
in the confinement phases, especially at $\beta$=0.99,
which is the same tendency as that reported in Ref.~\cite{Hart:1997vb}.
This feature may indicate the possible interrelation in compact QED
between the nonperturbative feature
and the appearance of large monopole clusters
which cover almost entire volume.

\subsection{Overlap fermion}

For the fermion action, we employ the overlap formalism~\cite{Neuberger:1997fp,Neuberger:1998wv}.
The overlap-Dirac operator $D$ is constructed as
\begin{equation}
D\equiv\rho[1+\gamma_5 {\rm sgn}(H_W)]
\equiv\rho\left[1+\gamma_5 \frac{H_W}{\sqrt{{H_W}^2}}\right]
\end{equation}
and realizes the exact chiral symmetry on a lattice satisfying the
Ginsparg-Wilson relation~\cite{Luscher:1998pq,Ginsparg:1981bj}
\begin{equation}
\gamma_5 D+D\gamma_ 5= \rho^{-1}D\gamma_5D.
\end{equation}
Here, $H_W\equiv \gamma_5 (D_W-\rho)$  is the hermitian Wilson-Dirac operator
defined with the standard Wilson-Dirac operator $D_W$.
The ``negative mass'' $\rho$ is chosen in the range of $0<\rho <2$,
which we set 1.6 throughout this paper.
We approximate the sign function ${\rm sgn}(H_W)$ 
by 150 degrees' Chebyshev polynomial~\cite{Giusti:2002sm},
treating ${\cal O}(200)$ lowest eigenmodes of $H_W$ exactly.
In the present analysis, we impose
the periodic boundary conditions in all the spatial direction
for the fermion fields,
whereas the anti-periodic boundary condition is imposed
in the temporal direction.
We compute lowest 50 eigenpairs for each $\beta$
implementing a restarted Arnoldi method~\cite{Morgan:2006}.
All the eigenvalues $\lambda_{\rm lat}$ of $D$,
which lie on a circle with the radius of $\rho$ in a complex plain,
are stereographically projected onto the imaginary axis
via M{\"o}bius transformation~\cite{Farchioni:1999se},
\begin{equation}
\lambda = \frac{\lambda_{\rm lat}}{1-\lambda_{\rm lat}/2\rho}.
\end{equation}

\section{Low-lying Dirac modes}
\label{diracmodes}

We briefly survey the properties of Dirac eigenmodes in this section.

\begin{figure}[h]
\begin{center}
\includegraphics[scale=0.27]{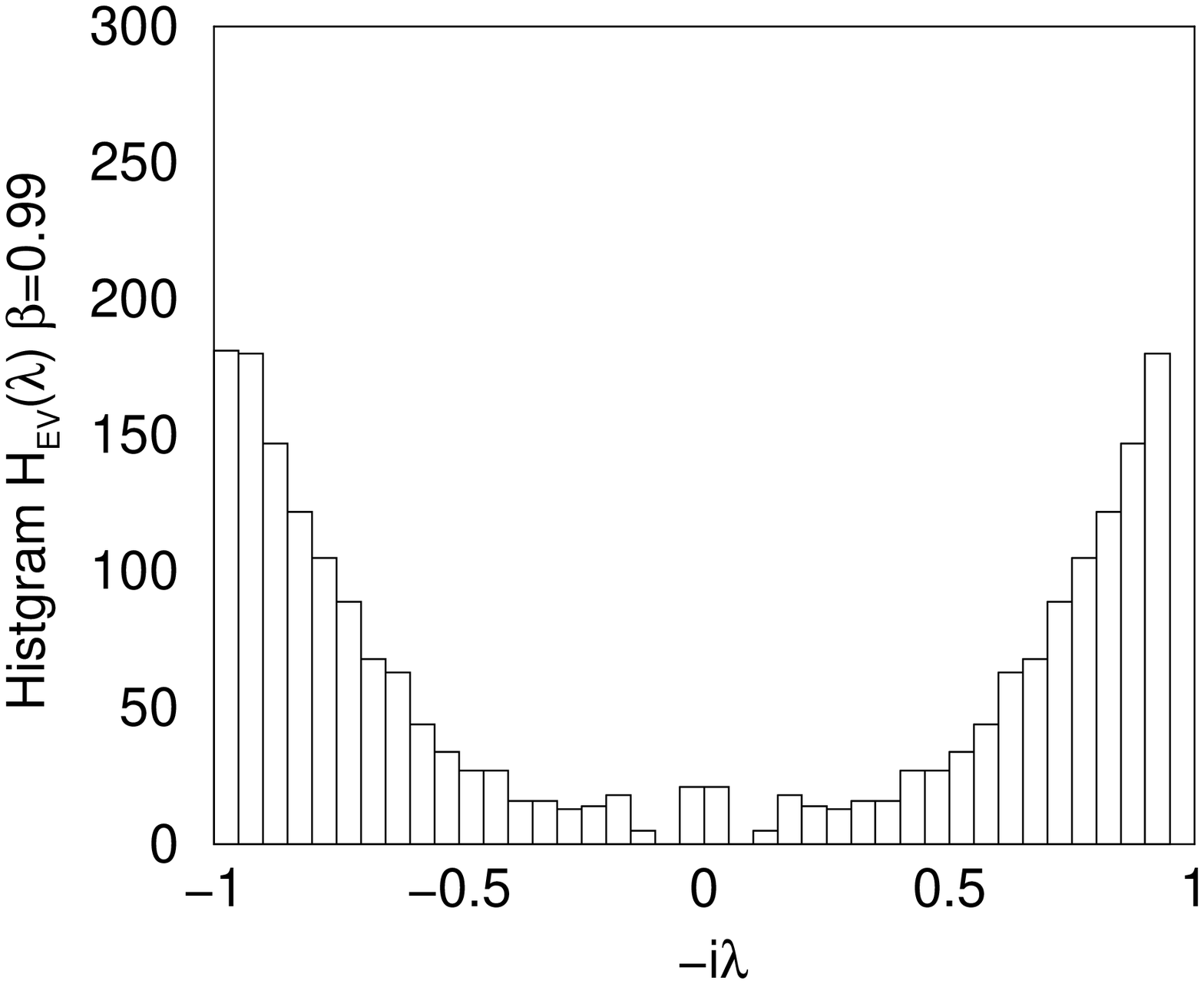}
\includegraphics[scale=0.27]{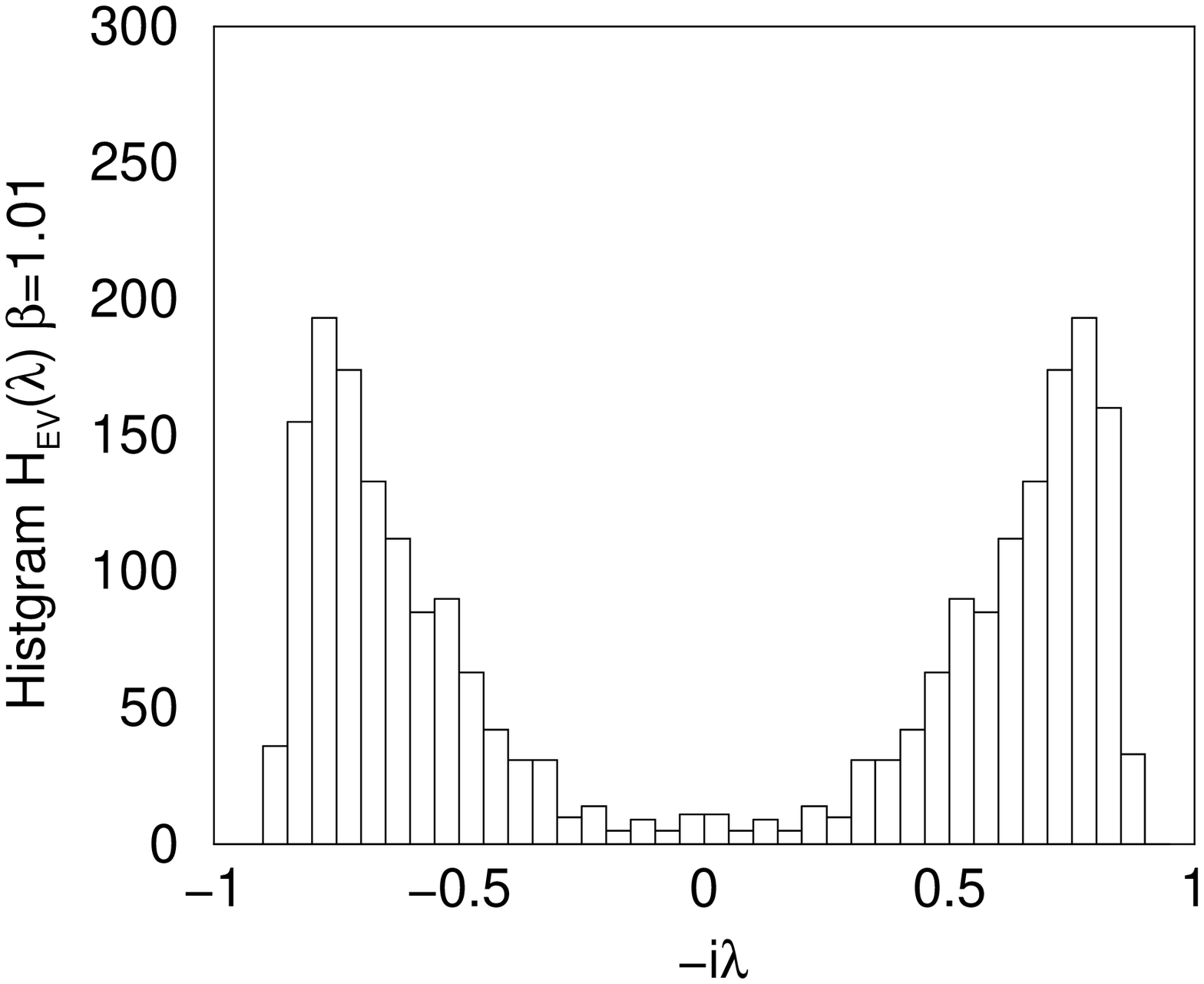}
\includegraphics[scale=0.27]{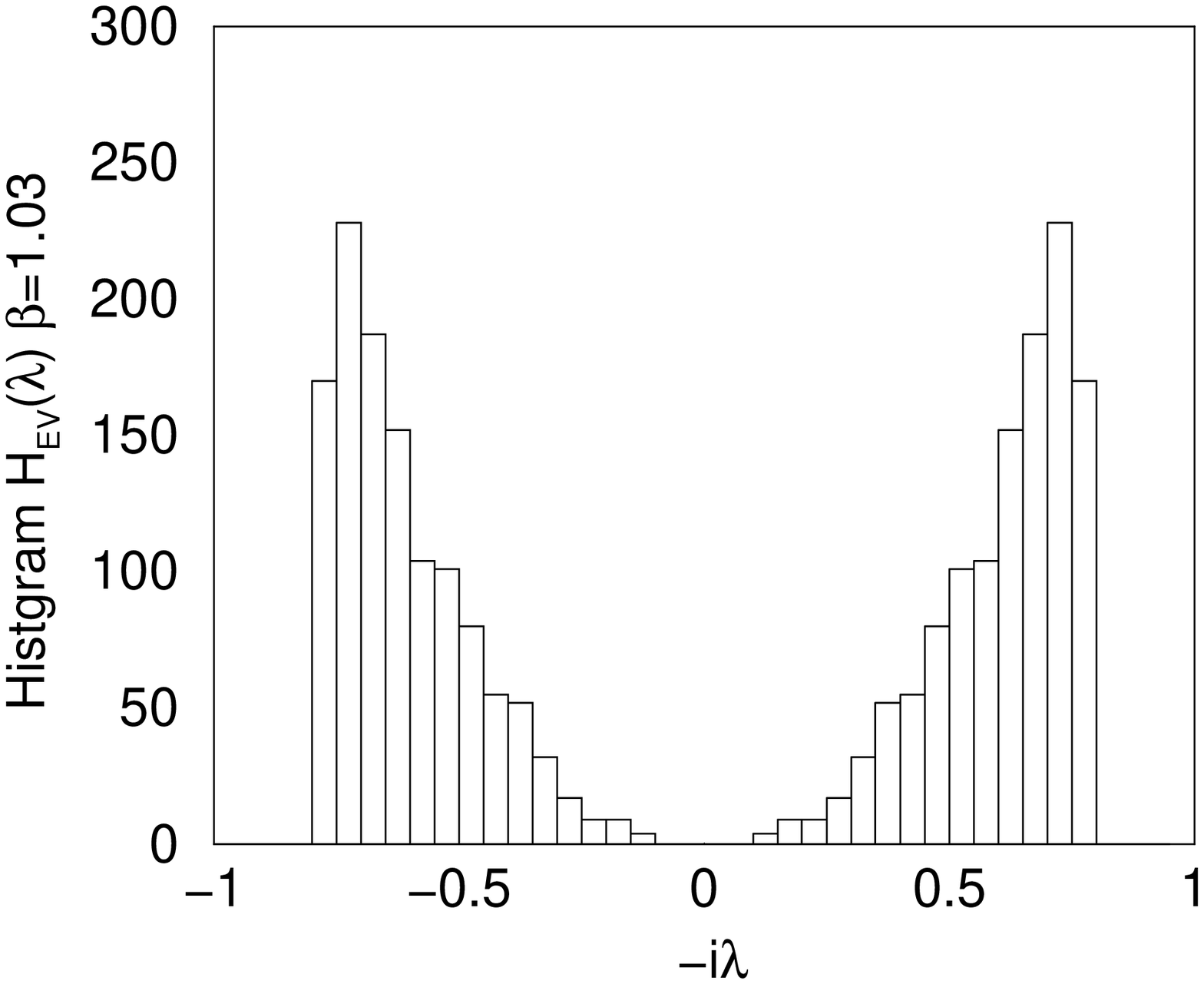}
\end{center}
\caption{\label{histgram01}
The histograms $H_{\rm ev}(\lambda)$ of the eigenvalues $\lambda$
of the overlap-Dirac operator $D$ are plotted for each $\beta$.
The horizontal axis denotes $-i\lambda$.
All the eigenvalues $\lambda_{\rm lat}$ lying 
on a circle in a complex plain
are stereographically projected onto the imaginary axis
via M{\"o}bius transformation,
$\lambda = \frac{\lambda_{\rm lat}}{1-\lambda_{\rm lat}/2\rho}$.
}
\end{figure}

We first show the histograms $H_{\rm ev}(\lambda)$ of 
Dirac eigenvalues in Fig.~\ref{histgram01},
where the horizontal axis denotes $-i\lambda$.
Since a non-zero eigenvalue is always accompanied
by its complex conjugate, the figure is symmetric 
about the $H_{\rm ev}$ axis.
We find exact zero modes in the confinement phase at $\beta$=0.99
and 1.01,
but on the other hand no zero mode is seen in the Coulomb phase
at $\beta$=1.03,
which is consistent with the previous works~\cite{Berg:2001nn,Drescher:2004st}.
The numbers of zero modes found in 48 gauge configurations at each
$\beta$ are listed in Table.~\ref{zeros}.
\begin{table}
\begin{center}
\begin{tabular}[h]{|c|c|c|c|}\hline
                  & $\nu=0$ & $\nu=1$ & $\nu=2$ \\ \hline
$\beta=0.99$      & 12      & 30      & 6       \\ \hline
$\beta=1.01$      & 32      & 12      & 4       \\ \hline
$\beta=1.03$      & 48      & 0       & 0       \\ \hline
\end{tabular}
\caption{\label{zeros}
The numbers of exact zero-modes found in 48 gauge configurations
are listed. The $i$-th column 
gives the number of configurations with 0, 1, 2 zero-mode(s), respectively.
}
\end{center}
\end{table}
The other prominent difference can be seen in the spectral density 
at the spectral origin.
At the strong coupling ($\beta$=0.99 and 1.01),
the density $\rho_{\rm ev} (\lambda)$ of near-zero modes
is rather dense, while
it rapidly goes to zero at $\beta$=1.03 as $\sim |\lambda_{\rm ev}|^3$.
These non-vanishing eigenvalue densities at $\lambda \sim 0$ are directly
connected to non-zero chiral condensate via Banks-Casher relation~\cite{Banks:1979yr} as
\begin{equation}
\langle\psi \bar \psi \rangle
\propto
\rho_{\rm ev}(0).
\end{equation}
Such dense spectral density 
is then considered as a signal of the broken chiral symmetry
in the strong-coupling compact QED.

We next extract the inverse participation ratio (IPR) for each
eigenmode at each $\beta$.
The IPR $I(\lambda)$ is defined as 
\begin{equation}
I(\lambda) = V\sum_x \rho_{\rm IPA}(x)^2,\quad
\rho_{\rm IPR}(x)\equiv \sum_{a,\alpha}|\psi_\lambda(x)|^2.
\end{equation}
Here, $V$ denotes the system volume and
$\psi_\lambda(x)$ is the eigenfunction 
associated with an eigenvalue $\lambda$
normalized as $\sum_x|\psi_\lambda(x)|^2=1$.
The Roman and Greek alphabets $a$ 
and $\alpha$ are the indices for a color and a spinor, respectively.
The density $\rho_{\rm IPR}(x)$
is obtained by locally summing up the absolute square of each component 
of an eigenfunction $\psi_\lambda(x)$ only over its color and spinor indices.
The IPR is unity when $\psi_\lambda(x)$ maximally spreads over the system
and equals to $V$ in the case when $\psi_\lambda(x)$ 
lives only on a single site,
reflecting the spatial distribution of the eigenfunction $\psi_\lambda(x)$.
\begin{figure}[h]
\begin{center}
\includegraphics[scale=0.27]{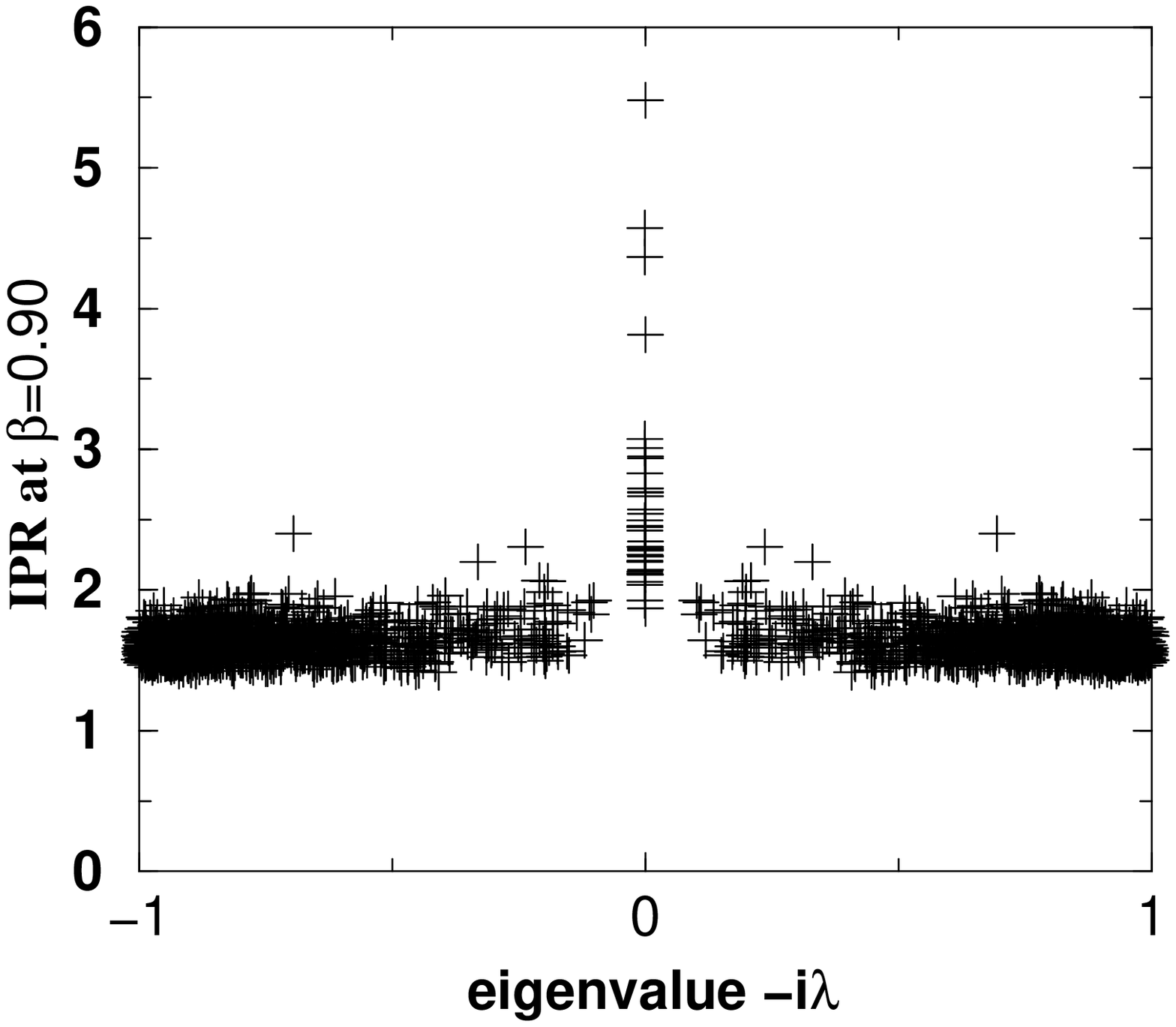}
\includegraphics[scale=0.27]{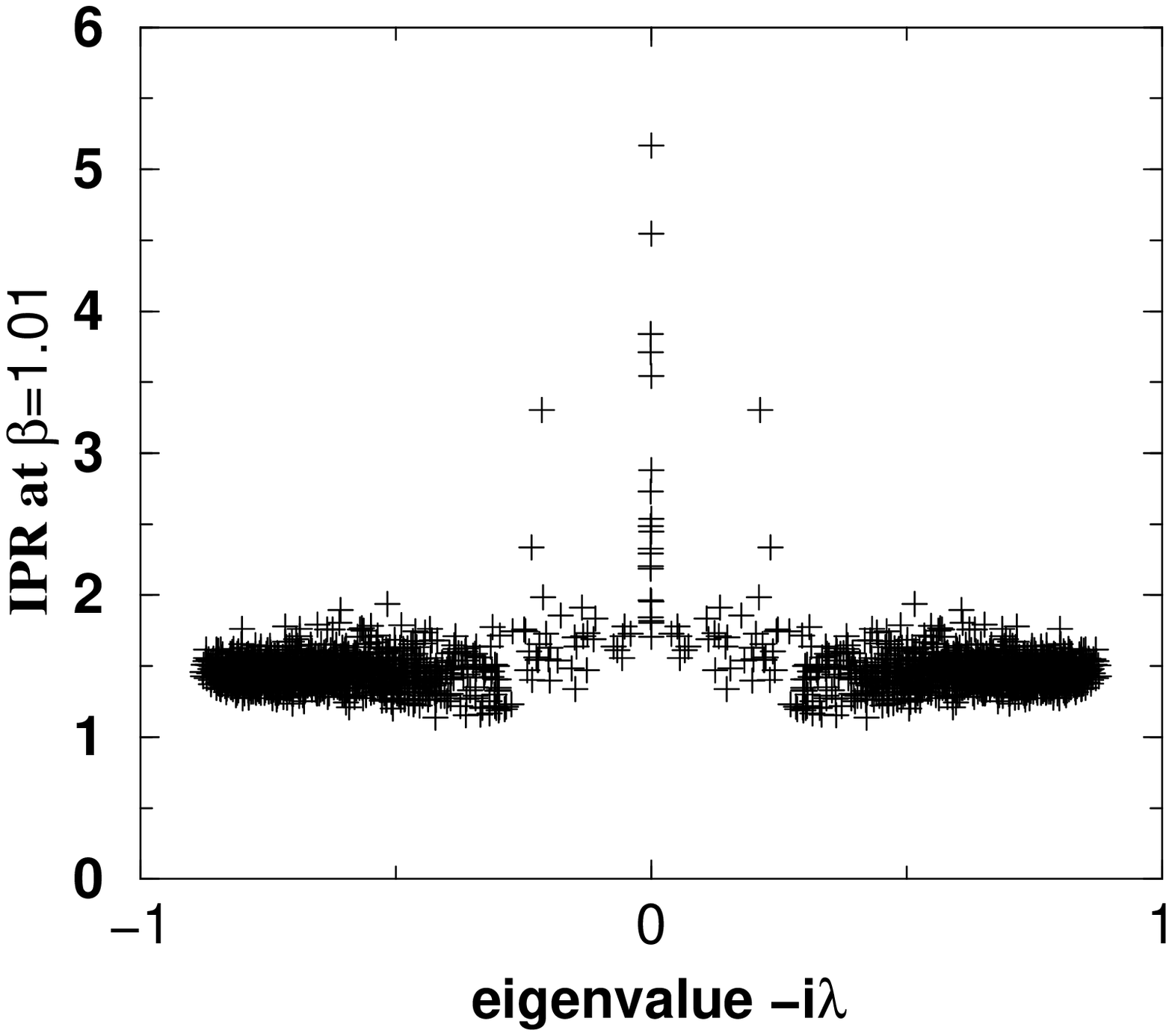}
\includegraphics[scale=0.27]{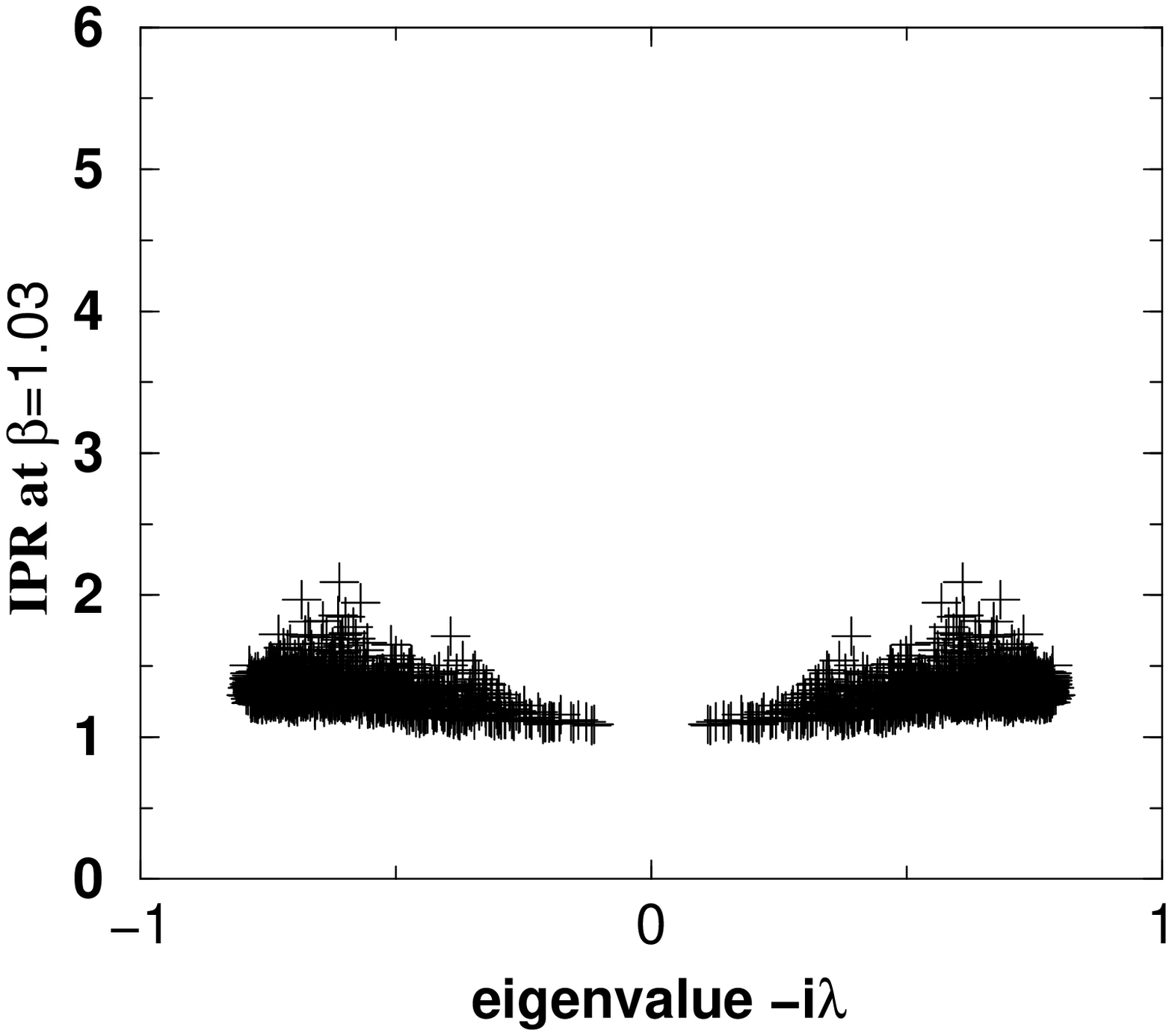}
\end{center}
\caption{\label{ipr}
The scattered plots of 
inverse participation ratios of low-lying Dirac modes
at $\beta$=0.99, 1.01 and 1.03.
The horizontal axis denotes $-i\lambda$,
the associated eigenvalue to each Dirac mode.
}
\end{figure}
As one can see in Fig.~\ref{ipr},
zero modes exhibit much larger IPRs than
the other low-lying modes (near-zero modes).
This tendency implies the stronger localization property
peculiar to zero modes in compact QED.
The IPRs for near-zero modes have $\beta$-dependence;
the IPRs at larger $\beta$ are 
smaller than those at smaller $\beta$ as a whole,
which implies that near-zero modes get delocalized
as we increase $\beta$.
This delocalization property of near-zero modes
was also found in Ref.~\cite{Drescher:2004st}.
Dirac modes in the weak coupling limit are expected to be plain waves
and to be completely delocalized.
The $\rho(\lambda)$ and the IPR at $\beta$=1.03 supports this expectation.

\section{Histograms of Dirac eigenmodes}
\label{analysis}

Our main concern in this paper
is a relationship between Dirac modes and monopoles.
Though the IPR surely reflects one aspect of a spatial distribution 
of an eigenfunction,
it can say almost nothing about the shape of an eigenmode.
The difficulty in such an analysis comes from 
the difficulty in a quantitative evaluation of the detail
of eigenfunction distributions.

\begin{figure}[h]
\begin{center}
\includegraphics[scale=0.26]{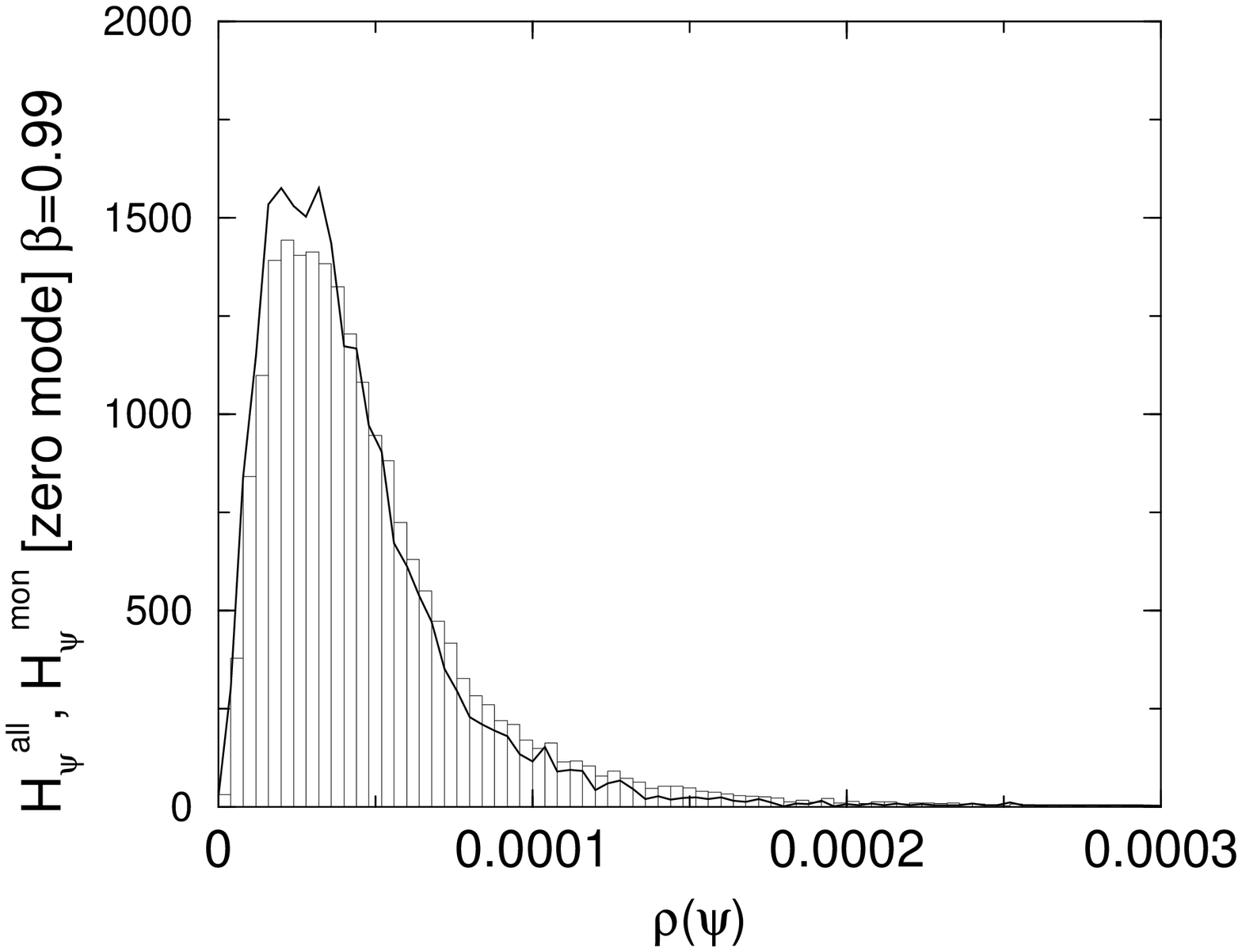}
\includegraphics[scale=0.26]{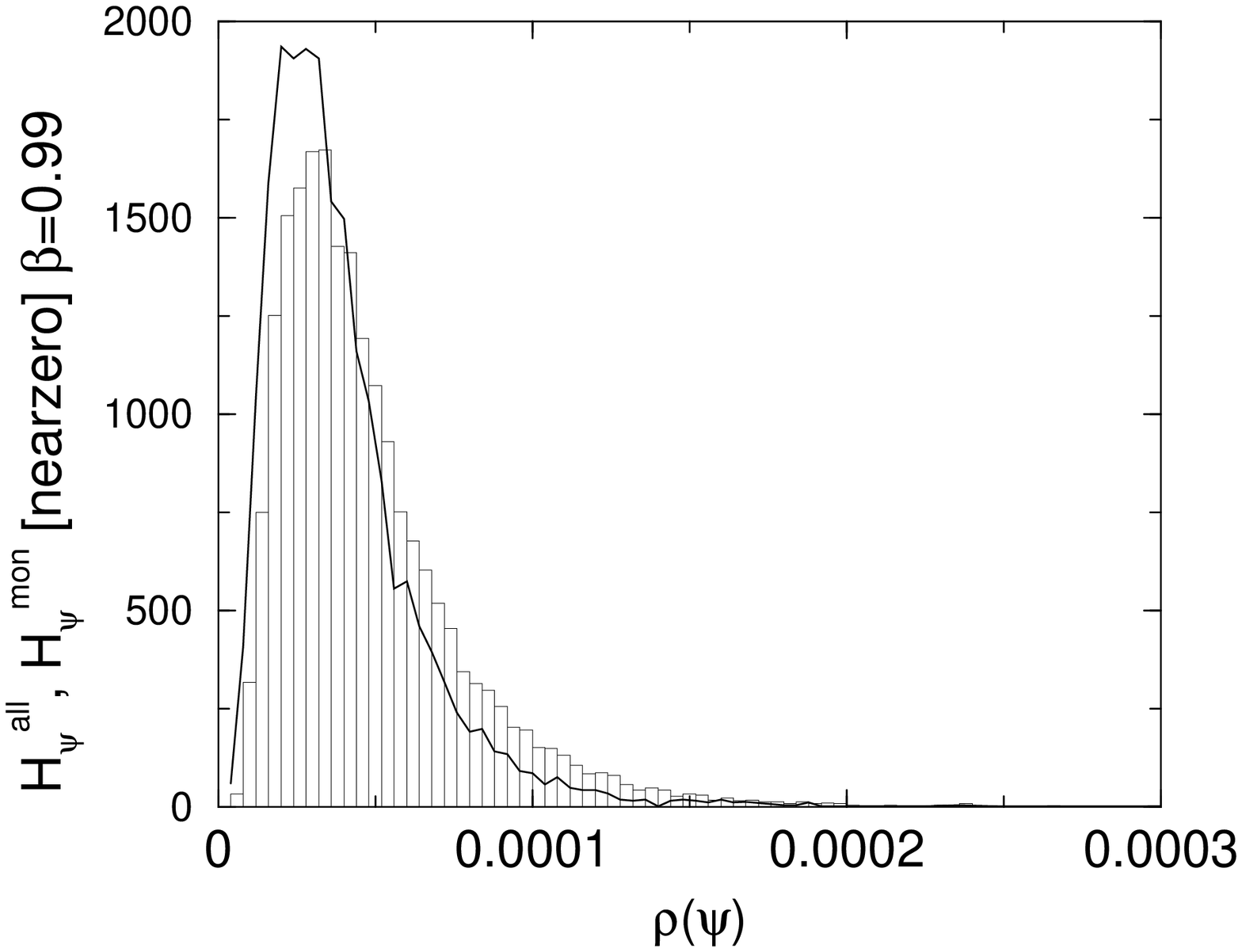}
\end{center}
\caption{\label{hgs01}
The density histograms $H^{\rm all}_{\rm \psi}$ and $H^{\rm mon}_{\rm \psi}$
at $\beta$=0.99
are plotted as the function of $\rho_\psi$.
The histograms $H^{\rm all}_{\rm \psi}$,
which are counted over all the site, are shown as open bars.
The histograms $H^{\rm mon}_{\rm \psi}$,
which are counted only on monopole currents are shown as solid lines.
}
\end{figure}

We first concentrate ourselves on the eigenfunction-density histograms
$H^{\rm all}_{\rm \psi}(\rho_{\psi})$.
The density $\rho_{\psi}\equiv \rho_{\rm IPR}$ here
is merely the same quantity as the IPR density $\rho_{\rm IPR}$
of a Dirac eigenfunction $\psi_\lambda(x)$.
The open bars in Fig.~\ref{hgs01} show the typical density histograms
$H^{\rm all}_{\rm \psi}$ counted at all the sites
for both zero and near-zero modes at $\beta$=0.99.
The histograms $H^{\rm all}_{\rm \psi}$
for zero modes have a relatively long tail at large $\rho_\psi$
and show a large height at $\rho_\psi\sim 0$
in contrast to those of near-zero modes,
which also exhibits the stronger localization property of zero modes.

We next investigate another type of histogram
$H^{\rm mon}_{\rm \psi}(\rho_\psi)$,
which is defined by counting only ``$\rho_\psi$ on monopoles''.
$H^{\rm mon}_{\rm \psi}$ is here normalized
as $\int_0^\infty H^{\rm mon}_{\rm \psi}(t)dt=
\int_0^\infty H^{\rm all}_{\rm \psi}(t)dt$.
Taking into account that monopoles are by definition located
at the intermediate points $\bar x\equiv x+(1/2,1/2,1/2,1/2)$,
$\rho_\psi$'s are simply redefined by averaging them around 
intermediate points $\bar x$ as
\begin{equation}
\rho_\psi\rightarrow\rho_\psi (\bar x)\equiv
\sum_{|x-\bar x|=1} \rho_\psi(x) / 2^4
\label{redef}
\end{equation}

In Fig.~\ref{hgs01}, we show the histograms 
$H^{\rm mon}_{\rm \psi}$ for both zero and near-zero modes
counted on the monopole world lines
at $\beta$=0.99 as solid lines.
At small $\rho_\psi$, we can find a remarkable difference
especially for near-zero modes.
At $\rho_\psi < 0.00005$, the histogram $H^{\rm mon}_{\rm \psi}$ 
counted only on monopoles takes a larger value
than $H^{\rm all}_{\rm \psi}$ counted over all the sites.
On the other hand, at $\rho_\psi > 0.00005$, $H^{\rm mon}_{\rm \psi}$ 
tends to be smaller than $H^{\rm all}_{\rm \psi}$.
This tendency indicates that monopoles ``run''
on the sites where the density $\rho_\psi$ of 
Dirac eigenfunctions is small. Or we can say that 
{\it near-zero Dirac modes are localized avoiding monopoles.}

We define and investigate the histogram ratios
$R_\psi(\rho_\psi)$
in order to evaluate the correlations in a semi-quantitative way;
\begin{eqnarray}
R_\psi(\rho_\psi)\equiv
\frac
{H^{\rm mon}_{\rm \psi}(\rho_\psi)}
{H^{\rm all}_{\rm \psi}(\rho_\psi)} ,
\end{eqnarray}
This quantity $R_\psi(\rho_\psi)$ equals to 1,
{\it if there is no correlation between the spatial fluctuations of Dirac
modes and monopoles}.
In the case when a positive (negative) correlation 
exists between the spatial fluctuations of Dirac modes and monopoles,
$R_\psi(\rho_\psi) > 1$ at smaller (larger) $\rho_\psi$
and
$R_\psi(\rho_\psi) < 1$ at large (smaller) $\rho_\psi$ hold.

\begin{figure}[h]
\begin{center}
\includegraphics[scale=0.26]{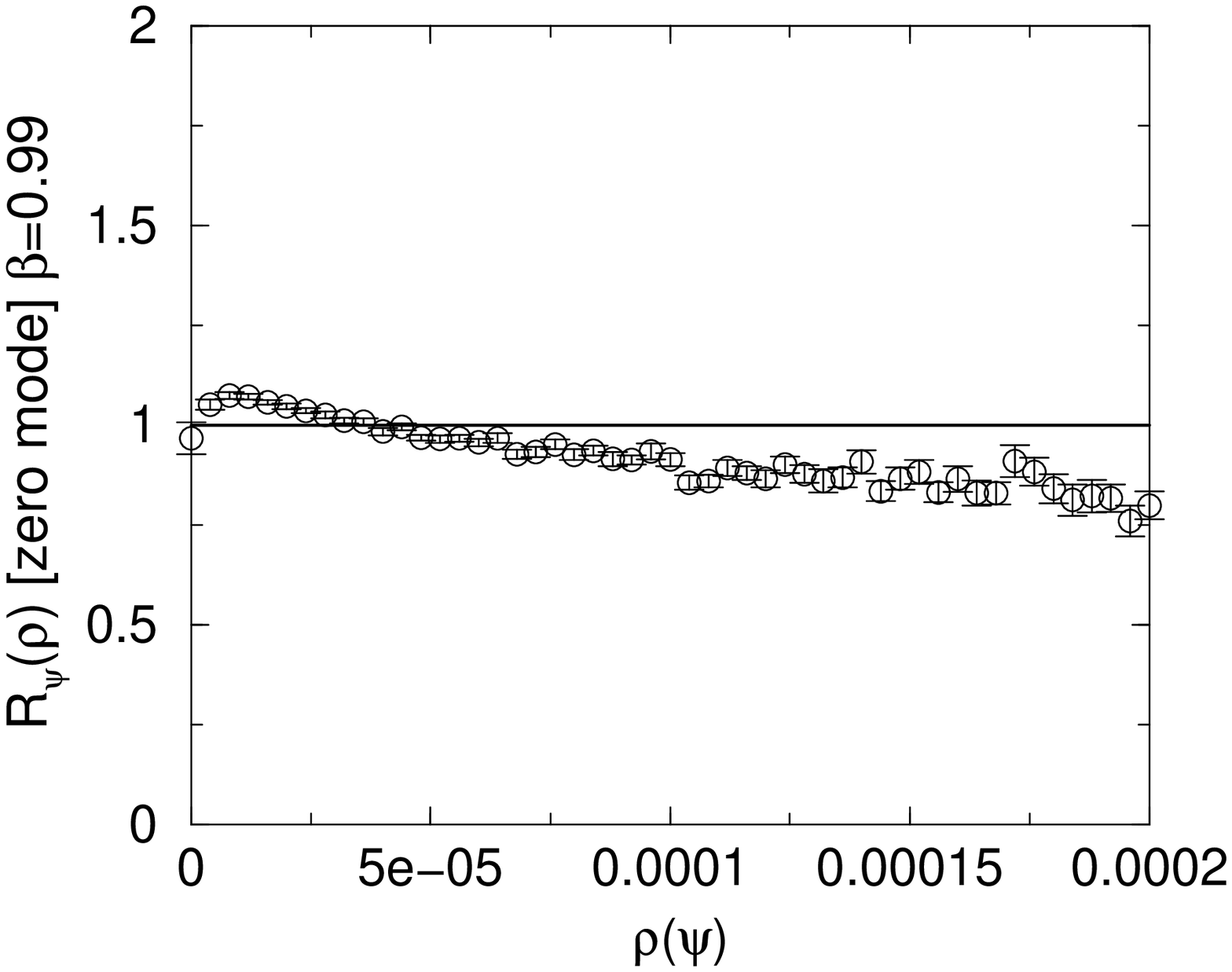}
\includegraphics[scale=0.26]{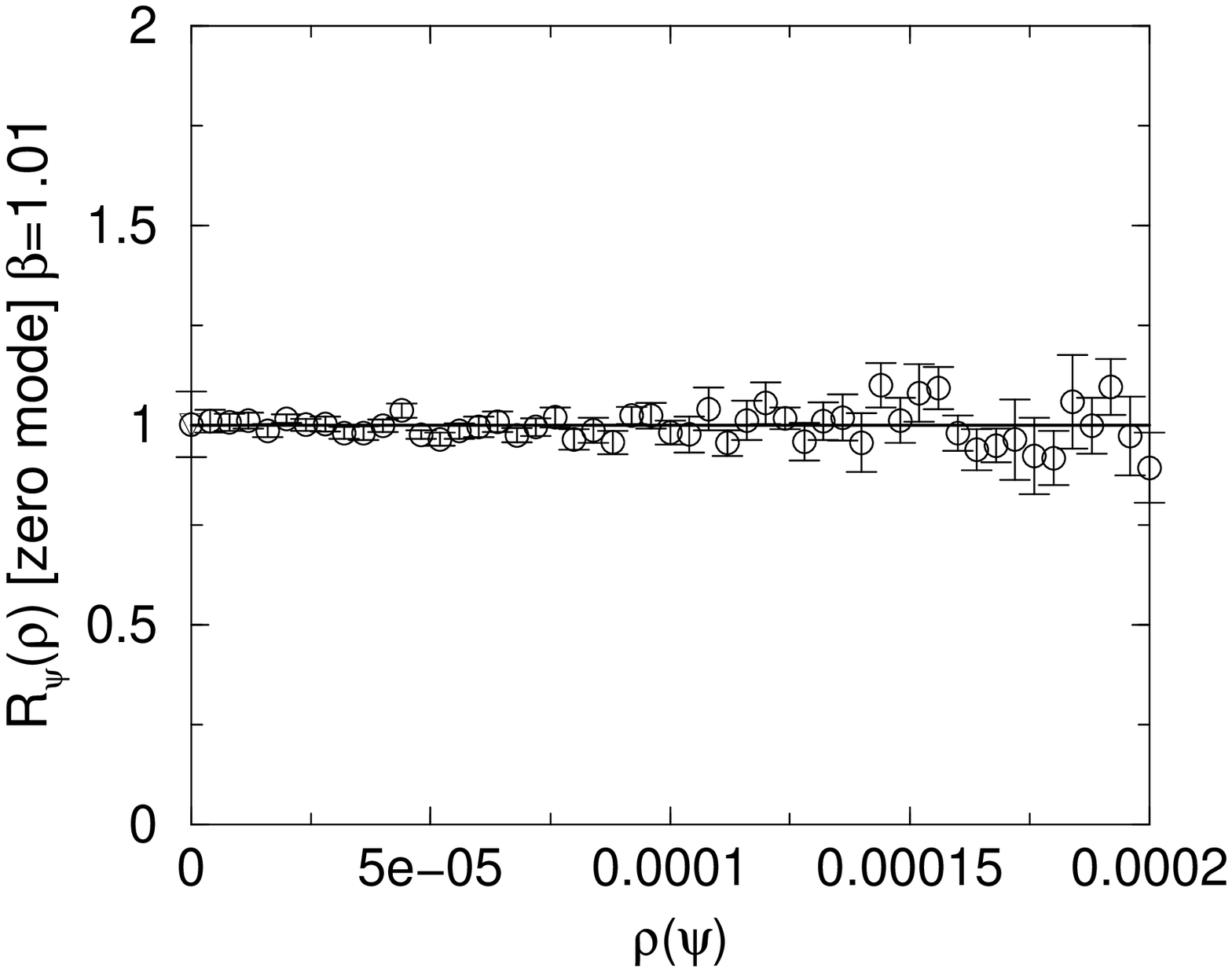} \\
\includegraphics[scale=0.26]{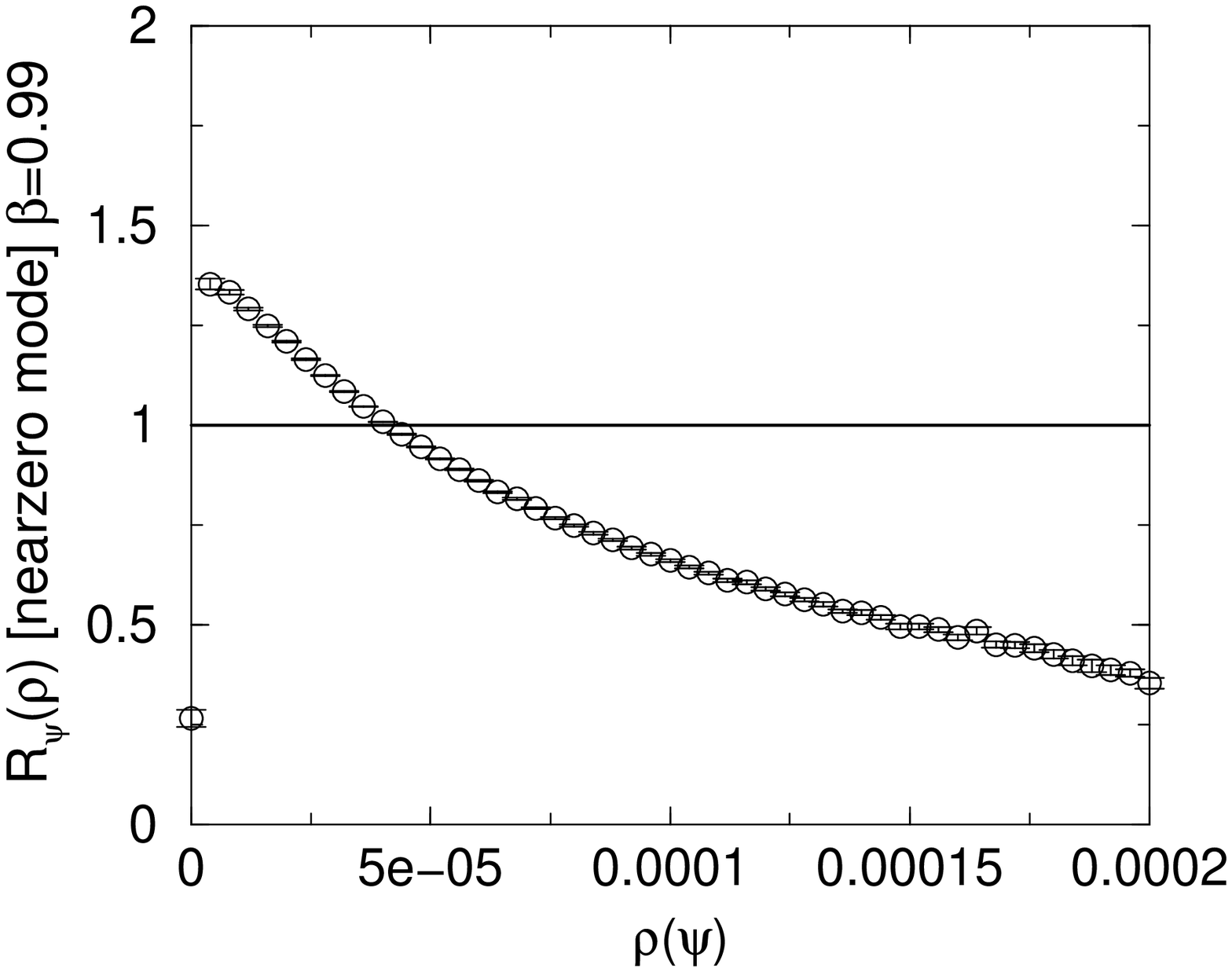}
\includegraphics[scale=0.26]{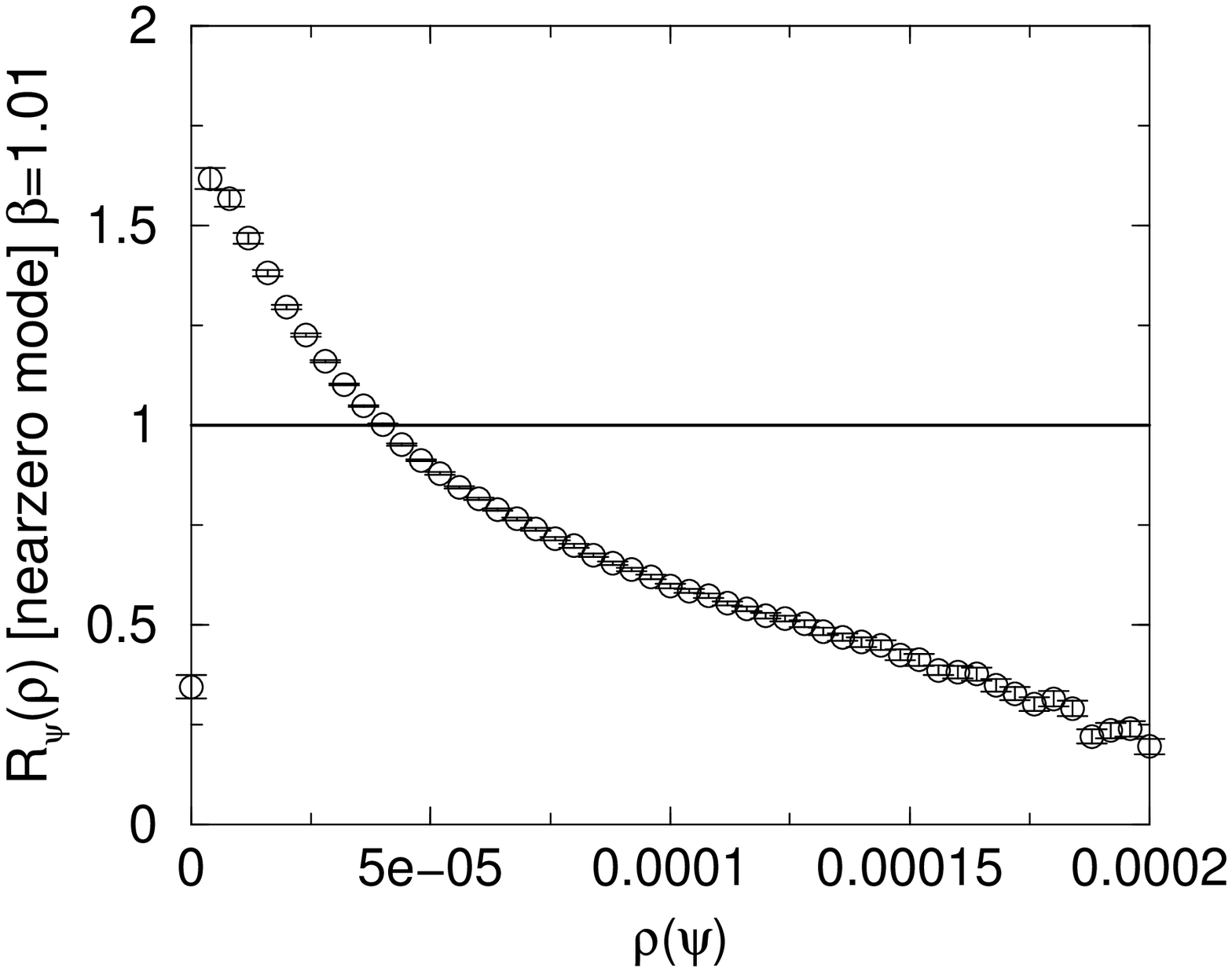}
\includegraphics[scale=0.26]{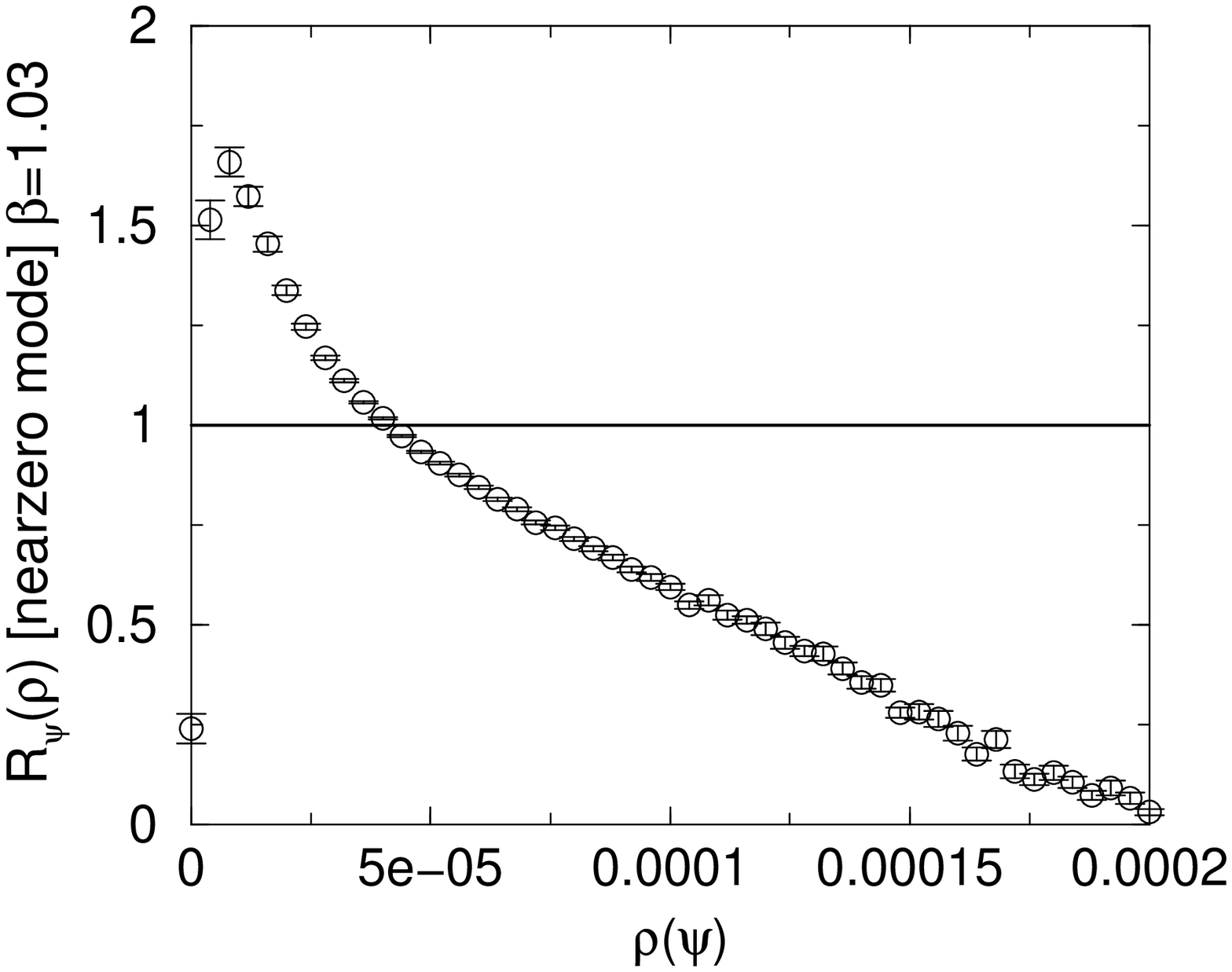}
\end{center}
\caption{\label{Ratios}
The histogram ratios $R_\psi(\rho_\psi)$
for zero (upper two panels) and near-zero modes (lower three panels)
at $\beta$=0.99, 1.01 and 1.03
are plotted in the range of $0 \leq \rho_\psi \leq 0.0002$.
We draw a line at $R_\psi(\rho_\psi)=1$ for reference.
}
\end{figure}

We here evaluate $R_\psi$ only with $\rho_\psi < 0.0002$,
because histograms $H_\psi$
are quite sparse at $\rho_\psi > 0.0002$ and 
this simple analysis cannot be reliable.
(For the $R_\psi(\rho_\psi)$ for near-zero modes,
we simply average $R$ over all the available near-zero modes.)
The ratios (to 1) of the squared norms of the Dirac modes
obtained by summing up only $\rho_\psi > 0.0002$
are 12(7)\%, 15(11)\% for zero modes at $\beta$=0.99, 1.01,
and 1.4(3)\%, 0.5(3)\%, 0.1(1)\% 
for near-zero modes at $\beta$=0.99, 1.01, 1.03, respectively.
As expected from the localization property of zero modes,
their ratios are larger than those of near-zero modes.
In spite of the truncation at $\rho_\psi \sim 0.0002$,
the analysis in the range of $\rho_\psi < 0.0002$
might be enough to grasp the nature of Dirac modes,
at least for near-zero modes, 
since $\rho_\psi > 0.0002$ are responsible 
only for a few \% of the total squared norms.
Fig.~\ref{Ratios} show
the histogram ratios $R_\psi(\rho_\psi)$
plotted in the range of $\rho_\psi < 0.0002$, respectively,
for both zero and near-zero modes at $\beta$=0.99, 1.01 and 1.03.

\subsection{Results for zero Dirac modes}
\label{Reszero}

The exact zero Dirac modes in compact QED have been studied
with regard to their relation to topological excitations~\cite{Berg:2001nn},
and to their spatial distributions~\cite{Drescher:2004st}.
The exact zero modes in compact QED can be found
only in the confinement phase~\cite{Berg:2001nn,Drescher:2004st}.
No detailed correlation between any 
of the topological phenomena and the zero-mode degeneracy
of the overlap-Dirac operator was found in Ref.~\cite{Berg:2001nn}.

While we have only a few dozen of zero modes in 48 gauge configurations
and the statistical errors are consequently rather large,
the spatial fluctuations of zero modes 
seem to have no remarkable correlation with monopole currents.
There are naively three possibilities:
1)The eigenfunctions fluctuate in a very complicated way
forming multifractal structures.
2)The zero-mode's fluctuation (especially $\rho_\psi < 0.0002$)
simply have little correlation with monopoles
and would be governed by some other possible objects.
3)The apparent correlations between zero modes and monopoles
appear only in the sector of $\rho_\psi > 0.0002$,
which we have discarded.

Even if the relationship between monopoles and zero modes is 
originally simple, the complicated monopole structure
could lead to the complex shape of zero modes,
which our present simple analysis may not detect.
(The monopole currents in fact occupy almost half of the 
total volume of $12^3\times 12=20736$
and their structures themselves are complicated.)
Moreover, the averaging in Eq.~\ref{redef} can be harmful in this case.
In order to verify this possibility 1),
it would be needed to compute multifractal dimensions of the modes
by analyzing the scaling of eigenfunction moments,
which is left for further study.
In the case of 2) and 3),
our present analysis can cast no more light on the nature of zero modes.
The authors in Ref.~\cite{Reinhardt:2002cm}
conjectured that Dirac zero modes mainly appear
at the intersections of vortexes. 
In such a rather complicated case, we need another analysis.

In any case, we would need more statistics and detailed analyses
in order to make definite conjectures upon zero modes.

\subsection{Results for near-zero Dirac modes}

As for near-zero Dirac modes,
one can see the outstanding {\it anti-correlation} in Fig.~\ref{Ratios}.
The histogram ratio $R_\psi(\rho_\psi)$ is clearly larger than 1
at small $\rho_\psi$ and less than 1 at large $\rho_\psi$
beyond the statistical errors.
Although the values of $R_\psi(\rho_\psi)$ themselves are slightly different
among three $\beta$'s, $\beta$=0.99, 1.01 and 1.03,
the bulk properties are surely the same,
which means the existence of the universal anti-correlations
between near-zero Dirac modes and monopole currents
around the critical coupling $\beta_c$.
The important point is the universality of the anti-correlation
between near-zero modes and monopoles.
This simple rule does not drastically change before/after the phase transition.
Near-zero modes are ``scattered'' by monopole clusters.

\section{Discussions and Speculations}
\label{discussions}

\subsection{Dirac eigenvalues and Chiral condensate}

As we have seen in Sec.~\ref{diracmodes},
Dirac eigenvalues have non-vanishing density $\rho_{\rm ev}(0)$
at the spectral origin in the chiral broken phase, 
and this non-vanishing value is directly connected
to the non-zero chiral condensate $\langle \bar \psi\psi\rangle$
via the Banks-Casher relation.
The non-vanishing $\rho_{\rm ev}(0)$
is generated by the ``repulsive force'' among the eigenvalues:
The eigenvalues of the overlap-Dirac operator
lying on the circumference of the circle in a complex plain
repel each other, and consequently near-zero eigenvalues
are pushed towards the spectral origin forming the non-zero $\rho_{\rm ev}(0)$.
Namely, the driving force of the chiral symmetry breaking
is the ``repulsive force'' among Dirac eigenvalues.
On the other hand, when $\rho_{\rm ev}(0)$ is zero, 
the repulsive force is considered to be weaker or zero.
The magnitude of this force can be revealed by level statistics.

\begin{figure}[h]
\begin{center}
\includegraphics[scale=0.27]{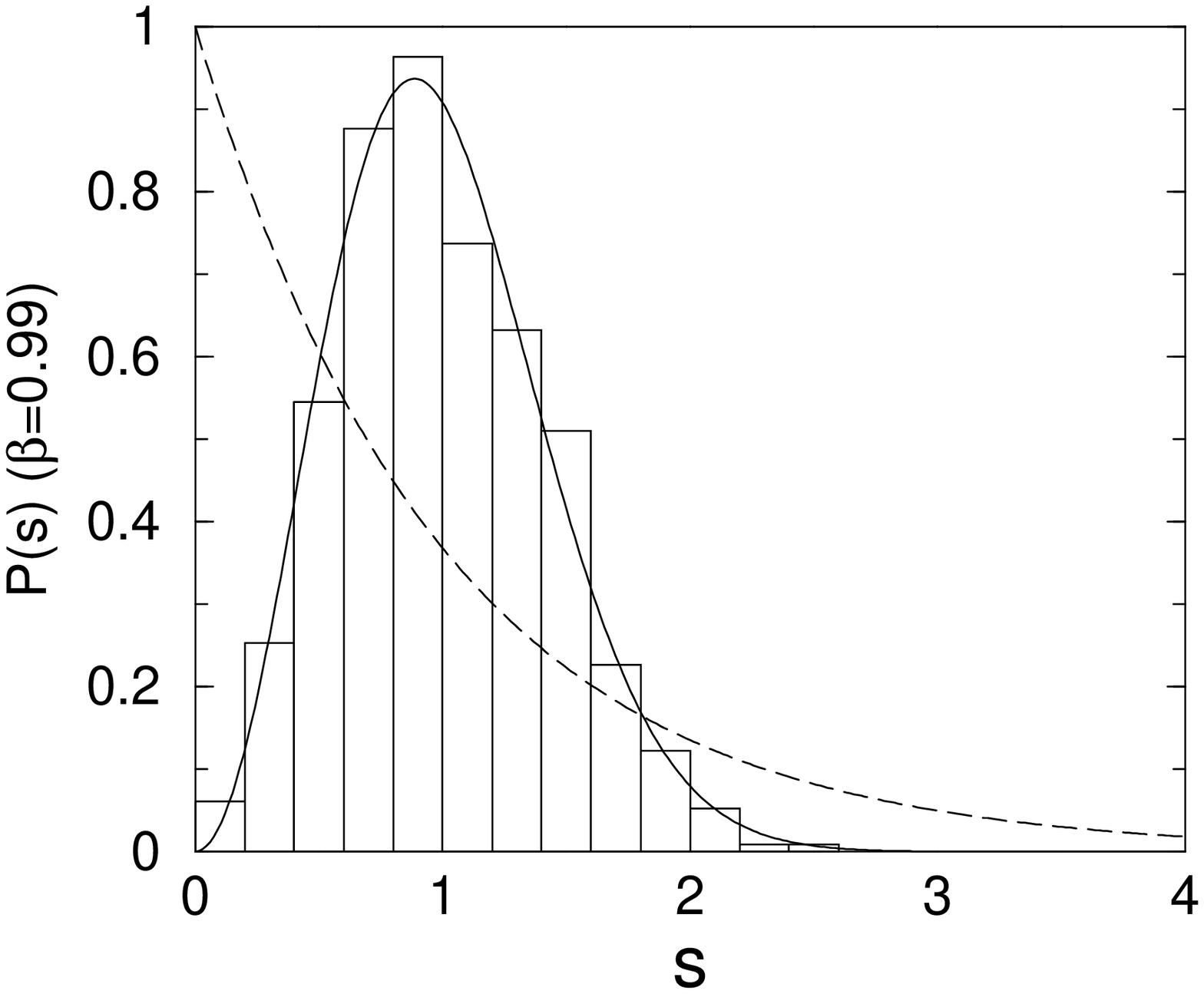}
\includegraphics[scale=0.27]{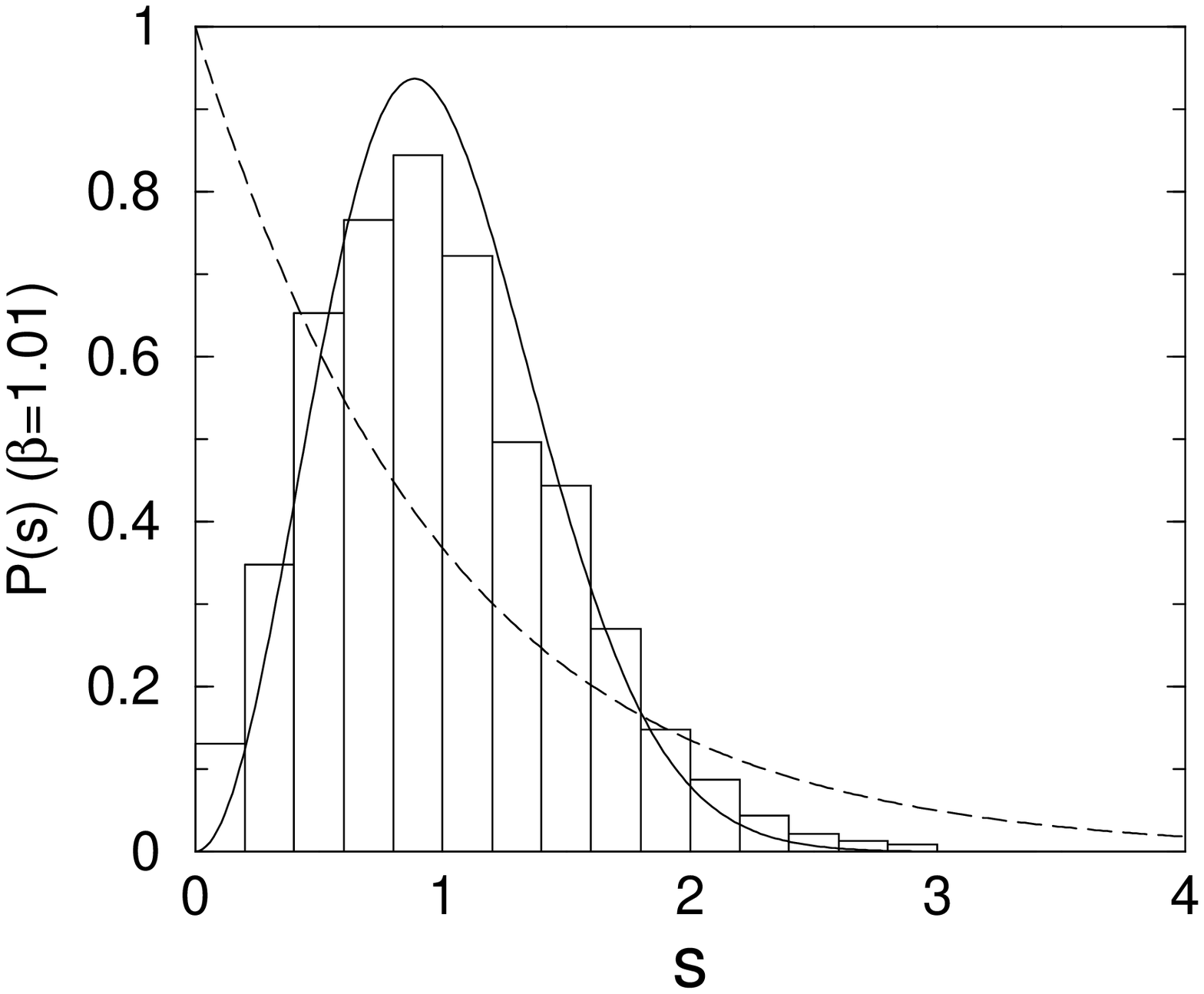}
\includegraphics[scale=0.27]{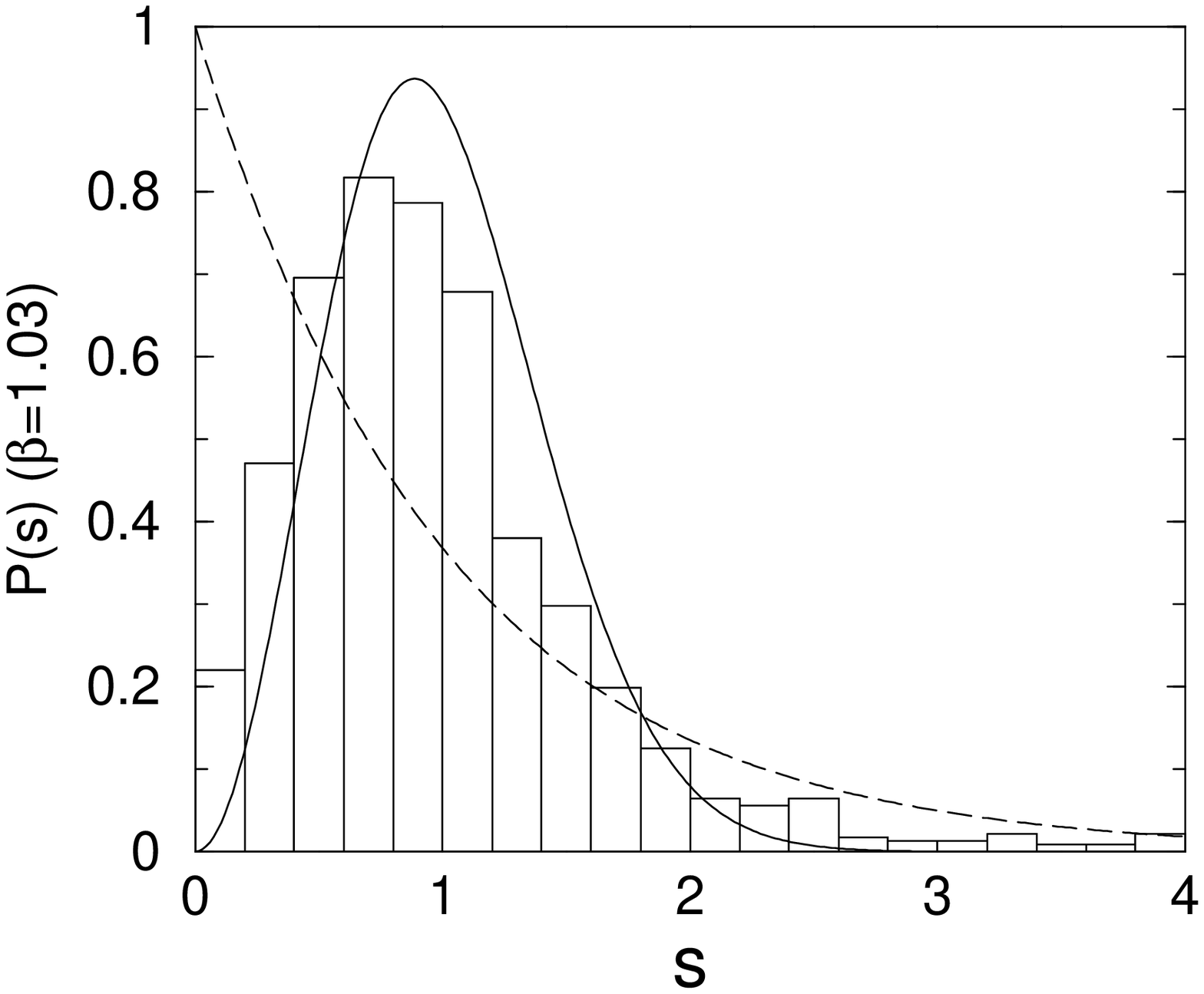}
\end{center}
\caption{\label{uls01}
The unfolded nearest-neighbor level spacing distributions $P_{\rm lat}(s)$
at $\beta=$0.99, 1.01 and 1.03.
The solid lines denote the Wigner distribution function
$P_{\rm Wig}(s)\equiv\frac{32}{\pi^2}s^2\exp (-\frac{4}{\pi}s^2)$
and the dashed lines the Poisson distribution function
$P_{\rm Poi}(s)=\exp (-s)$.
}
\end{figure}

We show in Fig.~\ref{uls01} the unfolded nearest-neighbor 
level spacing distributions $P_{\rm lat}(s)$
of the low-lying Dirac eigenvalues obtained at $\beta$=0.99, 1.01 and 1.03.
The solid lines denote the Wigner distribution function
$P_{\rm Wig}(s)\equiv\frac{32}{\pi^2}s^2\exp (-\frac{4}{\pi}s^2)$
and the dashed lines the Poisson distribution function
$P_{\rm Poi}(s)=\exp (-s)$.
The Wigner distribution function $P_{\rm Wig}(s)$
is a good approximation of the original distribution function
obtained by the random matrix theory (RMT) 
for the chiral unitary ensemble (chUE).
The Poisson distribution $P_{\rm Poi}(s)$ appears,
for example, in the system where eigenenergy levels
have {\it no correlation} with each other.

The level spacing distributions 
$P_{\rm lat}(s)$ at $\beta$=0.99 and 1.01
show the good coincidence with $P_{\rm Wig}(s)$.
The manifestation of the Wigner distribution,
at $\beta$=0.99 and 1.01 in the chiral broken phase, implies
the level repulsion among the low-lying eigenvalues,
which is consistent with the non-vanishing eigenvalue density 
$\rho_{\rm ev}(0)$ at the spectral origin.
On the other hand, at $\beta$=1.03 in the chiral restored phase,
$P_{\rm lat}(s)$ coincides with
neither the Wigner distribution $P_{\rm Wig}(s)$
nor the Poisson distribution $P_{\rm Pos}(s)$.
$P_{\rm lat}(s)$ at $\beta$=1.03
seems to be on the way from the Wigner to the Poisson distribution.
As a remarkable fact, you can find
the long tail of $P_{\rm lat}(s)$ stretching even at $s=4$.
This tendency seen at $\beta$=1.03
shows a weaker repulsion among the low-lying eigenvalues
than at $\beta$=0.99 and 1.01,
which leads to small or zero eigenvalue density $\rho_{\rm ev}(0)$ 
at the spectral origin.
The Wigner-Poisson transition in the nearest-neighbor level spacing
of the low-lying Dirac eigenvalues appears
at $\beta >1.01$, almost at the same time
as the confinement-deconfinement phase transition.

So far, we have mainly 
shown the numerical facts obtained in the present analysis.
In the next subsection,
we make a possible speculations about the chiral transition mechanism
in compact QED.

\subsection{Possible speculations}

What is the origin of a Wigner-Poisson transition
in the neighboring level spacing distributions of the low-lying
Dirac eigenmodes,
which is responsible for the chiral phase transition?
The Wigner/Poisson distributions
in neighboring level spacing distributions
can be also found in classically chaotic/regular systems.
Berry and Tabor~\cite{Berry:1977wk} drew the conclusion that
in a classically integrable system with more than one degree of freedom
the level spacing distribution of quantum spectra
obeys the Poisson distribution.
We can also find in Ref.~\cite{Bohigas:1983er}
the famous conjecture by Bohigas, Giannoni and Schmit
stating that the level statistics of a quantum system 
whose classical counterpart is chaotic 
do not show a dependence on the details of the dynamics
but depend only on the global symmetry of the system.
Level statistics in such a system coincide with those obtained 
by the random matrix theory with the same global symmetry.
(This conjecture is not always true.
A quantum kicked rotor does not satisfy the conjecture.)
The chaoticity or the complexity of the system 
leads to the Wigner distribution in level spacing.

The key player for the Wigner-Poisson transition 
in compact QED would be monopoles.
As was seen in Sec.~\ref{analysis},
there exist universal
anti-correlations between monopoles and near-zero Dirac eigenmodes.
In the system where large monopole clusters exist,
the near-zero modes are ``scattered'' by monopoles
in a very complicated way,
which results in the Wigner distribution of the neighboring level spacing.
When there exist only small monopole clusters in a system,
this complexity is much weakened
and the level spacing exhibits the Poisson-like distribution.
The level dynamics of low-lying Dirac eigenvalues
are simply controlled by monopoles' clustering-declustering feature.

Then the fact that the chiral symmetry breaking is
always accompanied by the charge confinement in compact QED
could be naturally understood from the microscopic viewpoint.
On one hand, the monopoles' clustering-declustering feature
is surely responsible for the confinement-deconfinement transition.
On the other hand,
the monopoles' clustering-declustering transition
gives rise to the change in the complexity of the system
and consequently in the level dynamics of low-lying Dirac eigenmodes,
which leads to the change in the eigenvalue density at the spectral origin,
or equivalently in the chiral condensate $\langle\bar\psi\psi\rangle$.

We finally note that the chaoticity and the level spacing distribution
in gauge theories were also investigated by several groups
~\cite{Berg:1998xv,Biro:1993qc,Biro:1997sq,Salasnich:1997cw,
Mukku:1996cn,Pullirsch:1998ke,Markum:2005ft}.

\section{Summary}
\label{summary}

We have studied the properties of low-lying Dirac modes
in quenched compact QED at $\beta$=0.99, 1.01 and 1.03,
employing $12^3\times 12$ lattices and the overlap formalism for the
fermion.

We have found several features worth noting:

\begin{itemize}
\item
The nearest-neighbor level spacing distribution of the Dirac eigenvalues
coincides with the Wigner distribution
at $\beta = 0.99,1.01< \beta_c$,
and it seems to be on the way from the Wigner to the Poisson
distribution at $\beta = 1.03 > \beta_c$,
which is consistent with 
the non-vanishing (vanishing) chiral condensate
at $\beta < \beta_c$ ($\beta > \beta_c$).
\item
The chiral phase transition in compact QED is 
very different from that in QCD at finite temperature.
The near-zero Dirac modes are much delocalized in the chiral restored
phase in compact QED,
whereas they are strongly localized in finite temperature QCD
exhibiting the Anderson-transition(AT) of the vacuum
~\cite{Anderson:1958vr,Garcia-Garcia:2006gr}.
\item
Near-zero modes have been found to have universal anti-correlations
with monopole world lines below/above the critical $\beta$.
\end{itemize}

The anti-correlation between monopoles and near-zero modes
indicates that near-zero modes are ``scattered'' by monopole currents
in a complicated way, which is considered as the origin
of the critical behaviors of low-lying Dirac eigenmodes.
The chiral phase transition in 4D compact QED
might be then associated with the complexity of the vacuum,
which is brought about 
by the clustering-declustering transition of monopoles.

\section*{acknowledgments}

The author thanks T.~Kunihiro, H.~Matsufuru, T.~Onogi and K.~Totsuka
for helpful comments,
and is supported by the Japan Society for
the Promotion of Science for Young Scientists. 
All the numerical calculations in the paper were carried out
on NEC SX-8 at the Yukawa Institute Computer Facility.

\end{document}